\documentclass[12pt]{article}

\usepackage{amsmath}
\usepackage{amssymb}
\usepackage{cite}

\def\nn{\nonumber}

\numberwithin{equation}{section}

\title{\bf \Large  Perfect fluid coupled to a solenoidal field which\\  enjoys the $\ell$-conformal Galilei symmetry}

\author{Timofei  Snegirev${}^{a}$\thanks{timofei.v.snegirev@tusur.ru}
\\[0.5cm]
\it{\small ${}^a$Laboratory of Applied Mathematics and Theoretical Physics,}\\
\it{\small Tomsk State University of Control Systems and Radioelectronics,}\\
\it{\small Lenin ave. 40, 634050 Tomsk, Russia}}

\date{}

\begin{document}

\maketitle

\begin{abstract}
A non-relativistic (Galilei-invariant) model of a perfect fluid
coupled to a solenoidal field in arbitrary spatial dimension is
considered. It contains an arbitrary parameter $\kappa$ and in the
particular case of $\kappa=1$ it describes a perfect fluid coupled to a
magnetic field. For a special value of $\kappa$, the theory admits the Schr\"{o}dinger symmetry group which is consistent
with the magnetic case in two spatial dimensions only.
Generalization to the case of the $\ell$-conformal Galilei group for
an arbitrary half-integer parameter $\ell$ is constructed.
\end{abstract}

\thispagestyle{empty}
\newpage
\setcounter{page}{1}

\section{Introduction}\label{Sec1}

Current interest in fluid dynamics with conformal symmetry
is caused by efforts to understand the hydrodynamic limit of the
AdS/CFT-correspondence (see review \cite{Ran09} and references
therein). In this limit, fluid mechanics can provide an effective
description of a strongly coupled quantum field theory. From the symmetry
standpoint, the main object here is the conventional conformal
algebra.

Application of the non-relativistic version of
AdS/CFT-correspondence \cite{Son08,BM08,NS2010} to strongly coupled
condensed matter systems generated a great deal of interest  in
non-relativistic conformal algebras. There are various conformal
extensions of the Galilei algebra, the most general of which is the
so-called $\ell$-conformal Galilei algebra \cite{Henkel,NOR97}. A
peculiar features of this algebra is that temporal and spatial
coordinates scale differently under the dilatation:  $t'=\lambda t$,
$x'_i=\lambda^\ell x_i$. In condensed matter physics,
$z=\frac{1}{\ell}$ is known as the dynamical critical exponent. The
algebra is finite-dimensional provided $\ell$ is a (half)integer
number. The cases $\ell=\frac12$ and $\ell=1$ are referred to as the
Schr\"{o}dinger algebra \cite{Nied72} and the conformal Galilei
algebra \cite{LSZ}, respectively. In recent years, dynamical
realizations of the $\ell$-conformal Galilei algebras attracted
considerable attention (see e.g. \cite{LSZ1}-\cite{CG} and
references therein).

As is known, the non-relativistic perfect fluid dynamics enjoys the
Schr\"{o}dinger symmetry provided a specific equation of state
is chosen \cite{RS00,JNPP04}. Quite recently, generalized
perfect fluid equations were formulated which enjoy the
$\ell$-conformal Galilei symmetry \cite{Gal22a} (see also
\cite{Gal22b,Sne23a}). In both cases, the fluid is considered as a
closed system in the absence of external forces. It is natural
to wonder whether the formalism in \cite{Gal22a} can be extended so as to
accommodate external forces. Of particular interest is the magnetic force. The goal of this
work is to study the case involving a solenoidal
field. To the best of our knowledge, this issue has not yet been discussed in
literature.

The paper is organized as follows. In the next section, the equations
of motion, the energy-momentum tensor and the Hamiltonian formulation are constructed
for a non-relativistic perfect fluid coupled to a solenoidal
field $B_i(t,x)$ in three spatial dimensions. They
contain an arbitrary parameter $\kappa$. In particular case of
$\kappa=1$ they describe the perfect magnetohydrodynamics. Symmetries
of the dynamical system under consideration are established and the
corresponding conserved charges are constructed. The Galilei
group is present for an arbitrary value of $\kappa$, while additional
conformal symmetries enlarging it to the Schr\"odinger group can be
realized for a special choise of $\kappa$ only. These results are generalized to
the case of an arbitrary number of spatial dimensions in Sect. \ref{Sec3}, where the
model of the perfect fluid coupled to antisymmetric solenoidal field
$F_{ij}(t,x)$ is considered. A generalization
corresponding to the $\ell$-conformal Galilei algebra for an arbitrary
half-integer value of the parameter $\ell$ is constructed as well. In the
concluding Sect. \ref{S4} we summarize our developments.

\section{Perfect fluid coupled to a vector field $B_i$}\label{Sec2}

In non-relativistic space-time parameterized by the coordinates $(t,x_i),\;i=1,...,d$, a fluid is
characterized by the density $\rho(t,x)$ and the velocity vector field
$\upsilon_i(t,x)$. Let us consider a model of the perfect fluid
coupled to a vector field $B_i(t,x)$ defined by the following
equations\footnote{Throughout the text we use the notations: $\partial_0=\frac{\partial}{\partial
t}$, $\partial_i=\frac{\partial}{\partial x_i}$, ${\cal
D}=\partial_0+\upsilon_i\partial_i$. Summation over repeated indices is understood.}
\begin{eqnarray}
&{\partial_0\rho}+ {\partial_i (\rho\upsilon_i)}=0,\label{EulEq1}
\\ &\rho{\cal
D}\upsilon_i=-{\partial_i p}-\kappa B_j\left({\partial_i
B_j}-{\partial_j B_i}\right),\label{EulEq2}
\\
&{\partial_0 B_i}=\kappa{\partial_j }\left(\upsilon_i
B_j-\upsilon_j B_i\right),\label{EulEq3} \\
& {\partial_i B_i}=0,\label{EulEq4}
\end{eqnarray}
where $p(t,x)$ is the pressure obeying the equation of state
$p=p(\rho)$ and $\kappa$ is an arbitrary parameter. For $\kappa=1$
and $d=3$ the above equations describe the perfect fluid coupled to
the magnetic field $B_i$ (see for example \cite{LL60}), which
include the continuity equation (\ref{EulEq1}), the Euler equation
in magnetic field (\ref{EulEq2}), the low-frequency version
of the Maxwell equations for a
medium of infinite conductivity (\ref{EulEq3}), (\ref{EulEq4}). The
last equation (\ref{EulEq4}) implies the vector field is solenoidal.

The equations above admit the Hamiltonian formulation
\begin{eqnarray}
\partial_0\rho=\{\rho,{ H}\},\quad\partial_0\upsilon_i=\{\upsilon_i,{ H}\},\quad
\partial_0{B}_i=\{B_i,{ H}\},
\end{eqnarray}
which relies upon the Hamiltonian
\begin{eqnarray}\label{Hamd3}
{ H}=\int
dx\left(\frac12\rho\upsilon_i\upsilon_i+V+\frac12B_iB_i\right)
\end{eqnarray}
and the Poisson brackets
\begin{eqnarray}\label{PBPF}
\{\rho(x),\upsilon_i(y)\}&=&- {\partial_i}\delta(x-y),\nn
\\
\{\upsilon_i(x),\upsilon_j(y)\}&=&\frac{1}{\rho}\left({\partial_i\upsilon_j}
-{\partial_j\upsilon_i}\right)\delta(x-y),\nn
\\
\{\upsilon_i(x),B_j(y)\}&=&-\frac{\kappa}{\rho}\left(B_j{\partial_i
} -\delta_{ij}B_k{\partial_k }\right)\delta(x-y),
\end{eqnarray}
in which derivatives on the right-hand sides are evaluated with respect to $x_i$.
It is assumed that the potential $V(\rho)$ entering (\ref{Hamd3}) is
related to the pressure via the Legendre transformation
\begin{eqnarray}\label{Leg}
p=\rho V'-V.
\end{eqnarray}

Note that, in order to verify the Jacobi identity,
properties of delta-functions should be used and total spatial
derivative terms must be disregarded. Within the context of magnetohydrodynamics ($\kappa=1$), the brackets
(\ref{PBPF}) were
formulated in \cite{MG80}.

The model under consideration enjoys the Galilei symmetry for an
arbitrary value of the parameter $\kappa$. The conserved energy coincides with the Hamiltonian ${
H}$, while the conserved momentum, the Galilei boost generator and the angular momentum generator read
\begin{eqnarray}\label{GalCh}
{ P}_i=\int dx\rho\upsilon_i,\quad { C}_i=P_it-\int dx\rho x_i,\quad
{ M}_{ij}=\int dx \rho (x_i\upsilon_j-x_j\upsilon_i).
\end{eqnarray}
They form the centrally extended Galilei algebra under the Poisson bracket
\begin{align}\label{GalA}
& \{{ H},{ C}_i\}={ P}_i, && \{{ M}_{ij},{ P}_k\}=-\delta_{k[i}{
P}_{j]},\nn
\\
& \{{ P}_i,{ C}_j\}=\delta_{ij}M, && \{{ M}_{ij},{
C}_k\}=-\delta_{k[i}{ C}_{j]},
\end{align}
where the central charge $M=\int dx\rho$ links to the mass of the fluid.

For a special choice of
$\kappa$ and $V$, one can construct additional conserved quantities. The conventional way to
prove this is to construct the non-relativistic energy-momentum
tensor $T^{\mu\nu},\;\mu=0,i$ associated with the equations
(\ref{EulEq1}--\ref{EulEq4}). The following energy density and flux
\begin{eqnarray}
T^{00}&=&\frac12\rho\upsilon_i\upsilon_i+V+\frac12B_iB_i,
\\
T^{i0}&=&\rho\upsilon_i(\frac12\upsilon_j\upsilon_j+V')+\kappa
 B_j(\upsilon_iB_j-\upsilon_jB_i)
\end{eqnarray}
satisfy
\begin{eqnarray}
{\partial_0 T^{00}}+{\partial_i T^{i0}}=0.
\end{eqnarray}
Similarly, the momentum density and stress tensor
\begin{eqnarray}
T^{0i}&=&\rho\upsilon_i,\quad
T^{ji}=\rho\upsilon_i\upsilon_j+\delta_{ij}p-\kappa(B_iB_j-\frac{1}{2}\delta_{ij}B^2)
\end{eqnarray}
obey
\begin{eqnarray}
{\partial_0 T^{0i}}+{\partial_j T^{ji}}=0.
\end{eqnarray}
Note that $T^{i0}\neq T^{0i}$ because the theory is not
Lorentz-invariant but $T^{ij}= T^{ji}$ as the theory is
invariant under spatial rotations.

If
\begin{eqnarray}
2T^{00}=\delta_{ij}T^{ij}
\end{eqnarray}
additional conformal symmetry exist \cite{JNPP04}. In our case this condition requires
\begin{eqnarray}\label{kappa}
V=\frac12 p d,\quad \kappa=\frac{2}{d-2}
\end{eqnarray}
and allows one to build conserved charges associated with the dilatation and special
conformal transformation
\begin{eqnarray}\label{ShrCh}
{ D}=t{ H}-\frac12\int dx \rho x_i\upsilon_i,\quad { K}=t^2{ H}-2t{
D}-\frac12\int dx \rho x_ix_i.
\end{eqnarray}
Together with ${ H}$ they form $so(2,1)$
subalgebra under the Poisson bracket
\begin{eqnarray}\label{so2,1}
\{H,D\}=H,\quad \{H,K\}=2D,\quad \{D,K\}=K.
\end{eqnarray}
The remaining structure relations read
\begin{eqnarray}
\{D,P_i\}=-\frac12P_i,\quad \{D,C_i\}=\frac12C_i,\quad
\{K,P_i\}=-C_i.
\end{eqnarray}
(\ref{GalA}) and (\ref{so2,1}) describe the
Schr\"{o}dinger algebra \cite{Nied72}.

Thus, we have formulated the perfect fluid equations coupled to the
solenoidal vector field $B_i$, which enjoy the Schr\"{o}dinger
symmetry, and constructed the corresponding Hamiltonian formulation.
The potential $V=\frac12  p d$ together with the relation
(\ref{Leg}) give the equation of state $p=\nu\rho^{1+\frac{2}{d}}$
($\nu$ is some constant), which is the same as in the absence of
external forces \cite{RS00}.

\section{Perfect fluid coupled to an antisymmetric field $F_{ij}$}\label{Sec3}

\subsection{The Schr\"{o}dinger symmetry}

Although the magnetic field ($\kappa=1$, $d=3$) does not fit the restriction (\ref{kappa}),
in an arbitrary spatial dimension, where one can introduce
an antisymmetric tensor $F_{ij}=-F_{ji}$ instead,
the Schr\"{o}dinger symmetry can be successfully accomodated. Below we discuss this issue in detail.

Consider a model governed by the equations
\begin{eqnarray}\label{PFEqd}
{\partial_0\rho}+ {\partial_i (\rho\upsilon_i)}=0,\quad \rho{\cal
D}\upsilon_i=-{\partial_i p}- \kappa F_{ai}{\partial_b F_{ab}},\quad
p=p(\rho),
\end{eqnarray}
\begin{eqnarray}\label{MaxEqd}
{\partial_0 F_{ij}}=\kappa\left(\partial_i( \upsilon_a
F_{ja})-\partial_j (\upsilon_a F_{ia})\right),\quad {\partial_i
F_{jk}}+{\partial_j F_{ki}}+{\partial_k F_{ij}}=0.
\end{eqnarray}
which reproduce (\ref{EulEq1})--(\ref{EulEq4}) in three spatial
dimensions provided $F_{ij}=\varepsilon_{ijk}B_k$\footnote{Here
$\varepsilon_{ijk}$ is the Levi-Cevita symbol in $R^3$.}.
Identifying $F_{ij}$ with the spatial (magnetic) components of the
electromagnetic field strength $F_{\mu\nu},\;\mu={0,i}$ one can take
these equations to define a perfect fluid coupled to the magnetic
field in arbitrary spatial dimension ($\kappa=1$). To the best of
our knowledge, Eqs. (\ref{PFEqd})--(\ref{MaxEqd}) have not yet been
discussed in the literature. The second equation in (\ref{MaxEqd})
implies the field $F_{ij}$ is solenoidal and results in the useful
identity $F_{ab}\partial_i F_{ab}=2F_{ab}\partial_aF_{ib}$.

The equations above can put into the Hamiltonian form
\begin{eqnarray}
\partial_0\rho=\{\rho,H\},\quad\partial_0\upsilon_i=\{\upsilon_i,H\},\quad
\partial_0{F}_{ij}=\{F_{ij},H\}
\end{eqnarray}
provided one introduces the Hamiltonian
\begin{eqnarray}\label{Hamd}
H&=&\int
dx\left(\frac12\rho\upsilon_i\upsilon_i+V+\frac14F_{ij}F_{ij}\right),\quad
p=\rho V'-V
\end{eqnarray}
and the Poisson brackets
\begin{eqnarray}\label{PBPFd}
\{\rho(x),\upsilon_i(y)\}&=&- {\partial_i}\delta(x-y),\nn
\\
\{\upsilon_i(x),\upsilon_j(y)\}&=&\frac{1}{\rho}\left({\partial_i\upsilon_j}
-{\partial_j\upsilon_i}\right)\delta(x-y),\nn
\\
\{\upsilon_i(x),F_{jk}(y)\}&=&\frac{\kappa}{\rho}\left(F_{ij}{\partial_k
}-F_{ik}{\partial_j }\right)\delta(x-y).
\end{eqnarray}

The non-relativistic energy-momentum tensor associated with
the equations (\ref{PFEqd}) and (\ref{MaxEqd}) reads
\begin{align}
& T^{00}=\frac12\rho\upsilon_i\upsilon_i+V+\frac{1}{4}F_{ij}F_{ij},
&&
T^{i0}=\rho\upsilon_i(\frac12\upsilon_j\upsilon_j+V')+\kappa\upsilon_aF_{ib}F_{ab},
\\
& T^{0i}=\rho\upsilon_i, &&
T^{ij}=\rho\upsilon_i\upsilon_j+\delta_{ij}p+\kappa(F_{ia}F_{ja}-\frac14\delta_{ij}F_{ab}F_{ab}).
\end{align}
It is conserved for an arbitrary value of the parameter
$\kappa$
\begin{eqnarray}
{\partial_0 T^{00}}+{\partial_i T^{i0}}=0,\quad{\partial_0
T^{0i}}+{\partial_j T^{ji}}=0.
\end{eqnarray}
Additional conformal symmetry entering the Schr\"{o}dinger group arises provided
$2T^{00}=\delta_{ij}T^{ij}$, which imposes the same restriction (\ref{kappa}) upon the
potential $V$ and determines the parameter
$\kappa$ as follows
$$
\kappa=\frac{2}{4-d}.
$$
Note that for $d=3$ one gets $\kappa=2$, which correctly reproduces
(\ref{kappa}) in the previous section. At the same time, the case
of a perfect fluid in a magnetic field (i.e. $\kappa=1$) admits the
Schr\"{o}dinger symmetry for $d=2$ only. Conserved charges
corresponding to the Schr\"{o}dinger group look like in the previous
section (\ref{GalCh}), (\ref{ShrCh}) with the energy $H$ defined by the
Hamiltonian (\ref{Hamd}).

\subsection{The $\ell$-conformal Galilei symmetry}

The $\ell$-conformal Galilei algebra \cite{NOR97} is the most
general non-relativistic conformal algebra which is characterized by
an arbitrary (half)-integer parameter $\ell$ and for $\ell=\frac12$ reproduces
the Schr\"{o}dinger algebra. In
addition to time translation $H$, dilatation $D$, special conformal
transformation  $K$, spatial rotations\footnote{In this section we
disregard spatial rotations.}, spatial translation
$C_i^{(0)}$ and Galilei boost $C_i^{(1)}$  it also contains constant
acceleration generators $C_i^{(k)}$, $k=2,...,2\ell$.
They obey the structure relations
\begin{align}\label{lCG}
& {[H,D]}=H, && {[H,C^{(k)}_i]}=kC^{(k-1)}_i,\nn
\\
& {[H,K]}=2D, && {[D,C^{(k)}_i]}=(k-\ell)C^{(k)}_i,\nn
\\
& {[D,K]}=K, && {[K,C^{(k)}_i]}={(k-2\ell)}C^{(k+1)}_i
\end{align}
and can be realized in a non-relativistic space-time $(t,x_i)$ by
the following operators \cite{NOR97}
\begin{eqnarray*}
H={\partial}_0,\quad D=t{\partial}_0+\ell x_i{\partial}_i,\quad
K=t^2{\partial}_0+2\ell tx_i{\partial}_i,\quad
C^{(k)}_i=t^k{\partial}_i.
\end{eqnarray*}

Consider the set of equations of motion
\begin{eqnarray}\label{PFEqdl}
{\partial_0\rho}+ {\partial_i (\rho\upsilon_i)}=0,\quad \rho{\cal
D}^{2\ell}\upsilon_i=-{\partial_i p}- \kappa F_{ai}{\partial_b
F_{ab}},\quad p=\nu\rho^{1+\frac{1}{\ell d}},
\end{eqnarray}
\begin{eqnarray}\label{MaxEqdl}
{\partial_0 F_{ij}}=\kappa\left(\partial_i( \upsilon_a
F_{ja})-\partial_j (\upsilon_a F_{ia})\right),\quad {\partial_i
F_{jk}}+{\partial_j F_{ki}}+{\partial_k F_{ij}}=0,
\end{eqnarray}
which reproduces (\ref{PFEqd}) and (\ref{MaxEqd}) for
$\ell=\frac12$. In the absence of $F_{ij}=0$, they hold invariant
under the action of the $\ell$-conformal Galilei group
\cite{Gal22a}. Below we will fix the value of $\kappa$ , for which
(\ref{PFEqdl}), (\ref{MaxEqdl}) accomodate the $\ell$-conformal
Galilei symmetry. To do this, we use the Hamiltonian formulation and
restrict ourselves to the case of a half-integer
$\ell=n+\frac12,\quad n=0,1,2,...$.

Introducing auxiliary fields $\upsilon^{(k)}_i$, $k=0,1,...,2n$
with $\upsilon^{(0)}_i=\upsilon_i$, one can rewrite the second equation
in (\ref{PFEqdl}) in the equivalent first order form
\begin{eqnarray}\label{HamFlEql}
{\cal D}\upsilon^{(k)}_i=\upsilon^{(k+1)}_i ,\quad \rho{\cal
D}\upsilon^{(2n)}_i=-{\partial_i p}- \kappa F_{ai}{\partial_b
F_{ab}}.
\end{eqnarray}
Then the Hamiltonian
\begin{eqnarray}\label{Hamdl}
H&=&\int dx
\left(\frac12\rho\sum_{k=0}^{2n}(-1)^k\upsilon^{(k)}_i\upsilon^{(2n-k)}_i+V+\frac14F_{ij}F_{ij}\right),\quad
V=\ell p d.
\end{eqnarray}
generates the dynamical equations as follows
\begin{eqnarray}
\partial_0\rho=\{\rho,H\},\quad\partial_0\upsilon_i^{(k)}=\{\upsilon_i^{(k)},H\},\quad
\partial_0{F}_{ij}=\{{F}_{ij},H\}
\end{eqnarray}
provided one introduces the Poisson brackets
\begin{eqnarray}\label{PBPFl}
\{\rho(x),\upsilon^{(k)}_i(y)\}&=&-\delta_{(k)(2n)}{\partial_i}\delta(x-y),\nn
\\
\{\upsilon^{(k)}_i(x),\upsilon^{(m)}_j(y)\}&=&\frac{1}{\rho}\left(\delta_{(k)(2n)}{\partial_i
\upsilon^{(m)}_j}- \delta_{(m)(2n)}{\partial_j
\upsilon^{(k)}_i}+(-1)^{k+1}\delta_{(k+m)(2n-1)}\delta_{ij}\right)\delta(x-y),\nn
\\
\{\upsilon^{(k)}_i(x),F_{jk}(y)\}&=&\frac{\kappa}{\rho}\delta_{(k)(2n)}\left(F_{ij}{\partial_k
}-F_{ik}{\partial_j }\right)\delta(x-y).
\end{eqnarray}
Here $\delta_{(k)(m)}$ is the Kronecker symbol. In the absence of
$F_{ij}=0$ the Hamiltonian and the brackets were formulated in a
recent paper \cite{Sne23a}.

Conserved energy is represented by the
Hamiltonian (\ref{Hamdl}), while conserved charges corresponding
to the dilatation, special conformal transformations and vector
generators read
\begin{eqnarray}
D&=&tH-\frac12\int
dx\rho\sum_{k=0}^{2n}(-1)^{k}(k+1)\upsilon_i^{(k)}\upsilon_i^{(2n-k-1)},\nn
\\
K&=&t^2H-2tD-\frac12\int
dx\rho\sum_{k=0}^{2n}(-1)^k\Big((n+1)(2n+1)-k(k+1)\Big)\upsilon^{(k-1)}_i\upsilon^{(2n-k-1)}_i,\nn
\\
C_i^{(k)}&=&\sum_{s=0}^k(-1)^s\frac{k!}{(k-s)!}t^{k-s}\int dx \rho
\upsilon^{(2n-s)}_i,\quad  k=0,...,2n
\end{eqnarray}
where $\upsilon^{(-1)}_i=x_i$. It turns out that $C_i^{(k)}$ are
conserved for an arbitrary value of $\kappa$, whereas the conservation of $D$ and $K$ imposes the restriction
$$
\kappa=\frac{2}{(2n+1)(4-d)}.
$$
Under the Poisson brackets the charges obey the algebra (\ref{lCG}),
which involves the central charge \cite{GM11}
\begin{eqnarray}
\{C_i^{(k)},C_j^{(m)}\}&=&(-1)^{k}k!m!\delta_{(k+m)(2n+1)}\delta_{ij}M,\quad
M=\int dx\rho.
\end{eqnarray}

\section{Conclusion}\label{S4}
To summarize, in this work we have formulated the equations of motion of a
non-relativistic perfect fluid coupled to a solenoidal field in arbitrary
spatial dimension. The
equations involve an arbitrary parameter
$\kappa$ and enjoy the Galilei symmetry. For
$\kappa=1$ and $d=3$ they reduce a perfect
fluid coupled to a magnetic field. It was demonstrated that the dynamical system admits
the Schr\"{o}dinger symmetry group for a specific
value of $\kappa$, which links to spatial dimension. The corresponding Hamiltonian formulation was built and
the full set of conserved charges, which obey the Schr\"{o}dinger algebra under the Poisson bracket,
was presented.

Equations describing a perfect fluid coupled to
a solenoidal field admitting the $\ell$-conformal Galilei symmetry were proposed as well.
For an arbitrary half-integer $\ell$,  both the Hamiltonian formulation and the full set of
conserved charges were constructed. It was shown that under the Poisson bracket the latter
form the centrally extended $\ell$-conformal Galilei algebra.

\section*{Acknowledgements}
This work was supported by
the Russian Science Foundation, grant No 23-11-00002.

\end{document}